# Soft Phonon Modes and Diffuse Scattering in Pb(In$_{1/2}$Nb$_{1/2}$)O$_3$-Pb(Mg$_{1/3}$Nb$_{2/3}$)O$_3$-PbTiO$_3$ Relaxor Ferroelectrics

Qian Li[1,a], Sergey Danilkin[2], Guochu Deng[2], Jian Wang[1], Zhengrong Li[3], Ray L. Withers[1], Zhuo Xu[3] and Yun Liu[1,b]

[1]Research School of Chemistry, The Australian National University, ACT 0200, Australia

[2]Bragg Institute, Australian Nuclear Science and Technology Organisation (ANSTO), NSW 2232, Australia

[3]Electronic Materials Research Laboratory, Xi'an Jiaotong University, Shaanxi 710049, PR China

**Abstract**

Single crystals of a ternary relaxor ferroelectric system, 0.29Pb(In$_{1/2}$Nb$_{1/2}$)O$_3$-0.45Pb(Mg$_{1/3}$Nb$_{2/3}$)O$_3$-0.26PbTiO$_3$, have been studied using triple-axis based elastic and inelastic neutron scattering. Elastic diffuse scattering confirms the presence of polar nano-regions (PNR's) in this system. The PNR's emerge at the Burns temperature, $T_B$ ~630 K and then grow continuously in population and correlation size as the crystal cools down to 100 K. At 300 K, characteristic "butterfly" and ellipsoid shaped diffuse scattering patterns are observed on the *HK0* reciprocal space plane. Electrical poling along the [110] direction produces a marked asymmetry in the diffuse scattering patterns, with the parallel-to-the-field components enhanced while the perpendicular-to-the-field components suppressed. Several low energy phonon branches along the [001] and [111] directions were studied. Most significantly, the PNR-acoustic phonon coupling is confirmed for the [110] transverse acoustic (TA) phonons polarized along the [1-10] real space direction and the [100] TA phonons. This coupling appears to be anisotropic and correlated with the PNRs' a, b) Email: liq1@ornl.gov; yliu@rsc.anu.edu.au

distribution, and also affected by the relative length scales of the PNRs and phonon wave vectors. The well-known "waterfall" phenomenon is observed on the [001] and [110] transverse optical (TO) branches, near the zone center. The optical phonon measurements also reveal a lowest-energy, zone center soft TO mode, whose squared phonon energy increase linearly with decreasing temperature below the $T_\mathrm{B}$.

## 1. Introduction

Lead-based complex perovskite (chemical formula: Pb($B^1B^2$)$O_3$) relaxor ferroelectrics have been studied for over fifty years. Especially, the related studies soared during last two decades after the findings of highly exploitable, giant piezoelectricity in relaxor single crystals of Pb(Mg$_{1/3}$Nb$_{2/3}$)O$_3$-PbTiO$_3$ (PMN-PT) and Pb(Zn$_{1/3}$Nb$_{2/3}$)O$_3$-PbTiO$_3$ (PZN-PT).[1,2] Many of these studies were targeted to understanding the microscopic origin of the piezoelectric properties of relaxor materials as well as their dielectric relaxation behavior, using a variety of structural and spectroscopic techniques with different length and time scales probing capabilities. Among them, neutron scattering arguably has revealed critical insights due to its broad coverage of the momentum ($Q$)-energy ($E$) phase space.[3,4]

In terms of the average structures, relaxors in general are ordinary. For example, pure PMN has a pseudo-cubic ($C$) average structure that undergoes no structure transitions (thus no macroscopic polarization occurs) within the broad studied temperature range unless a sufficiently high electrical field is applied; PZN shows somewhat more complex behavior and sometimes appears as a rhombohedral ($R$) ferroelectric phase at room temperature. Introducing tetragonal ($T$) normal ferroelectric PT into PMN (or PZN, etc.) leads to a gradual transition from the pseudo-$C/R$ structure to the $T$ structure accompanied by a loss of macroscopic relaxor properties, and a so-called morphotropic phase boundary (MPB) exists in between the $R/T$ phases. On the other hand, it is well recognized that the local structures, primarily those of a polar correlation nature, play a fundamental role in generating the relaxor properties. The local polar correlations in relaxors are traditionally termed as polar nanoregions (PNR's) and believed to emerge below the Burns temperature $T_B$, first revealed by the observation of a departure from the linear temperature dependence of optical refractive index in PMN.[5] Neutron diffuse scattering has been intensively utilized to probe the local structures of relaxors and a wealth of experimental observations accumulate to date for a number of material systems. These observations include the reciprocal space distribution patterns of the diffuse scattering and its dependence on variables of composition, temperature, applied electric field and so on.[6-10] There are also some measurements addressing the wide range dynamics of the PNR-related diffuse scattering thus directly linking these local degrees of freedom with the macroscopic dielectric behavior.[11] Nevertheless, despite these solid experimental evidence in support of the existence of PNR's, the exact microscopic picture of PNR remains inconclusive among several existing interpretations.[12]

Lattice dynamics studies provide another approach to understanding relaxors. The early phonon measurements were emphasized to search for the zone-center soft mode of relaxors,[13-15] similar to that established in the canonical displacive system of PT.[16,17] It was found that the near zone-center transverse optical (TO) modes of PZN-0.08PT and PMN soften when approaching the $T_B$ from high temperature and become overdamped below the $T_B$. These studies also uncovered a so-called "waterfall" phenomenon that constant energy scans of the low-lying TO modes suggest an abrupt dispersion curve starting at certain wave vectors $q_{wf}$. Initially, the waterfall phenomena was postulated to be indicative of a coupling between the PNR and TO modes in the relaxors,[15] but later similar results observed on non-relaxor systems, *e.g.* PMN-0.6PT, unjustified such a postulation making this phenomena still a controversial issue now.[18] The transverse acoustic (TA) modes, on the other hand, have been convincingly shown to couple with the PNR's for PMN and PZN-0.045PT single crystals.[19-21] The TA-PNR coupling strongly influences the elastic properties of relaxor crystals and is very likely to be the crucial mechanism for the giant piezoelectricity of relaxor-PT single crystals.

In this work, we extended the neutron scattering studies to a ternary relaxor, Pb(In$_{1/2}$Nb$_{1/2}$)O$_3$-Pb(Mg$_{1/3}$Nb$_{2/3}$)O$_3$-PbTiO$_3$ (PIN-PMN-PT), with a composition near the MPB of the system. Compared to PMN-PT, this system shows improved electrical properties, such as higher Curie temperature $T_C$ and coercive fields, making them more favorable for applications.[2] PIN itself exhibits a complex phase constitution (relaxor, ferroelectric or antiferroelectric) depending on the *B*-site ordering extent. The lattice dynamics of PIN in relation to the *B*-site ion randomness has been studied by Ohwada *et al* using inelastic neutron and X-ray scattering.[22,23] Mixing PIN into PMN-PT results in a further charge-frustrated state in which the *B*-site valence varies from +2 to +5. Thus, it is of some considerable interest to carry out a phonon study on PIN-PMN-PT to examine the lattice dynamic behavior previously observed in other relaxors and look for the features potentially specific to this system. On the other hand, the effects of electrical fields applied along the [110] direction (in pseudo-*C* settings) on the local structure and phonon behavior of relaxors have been rarely addressed thus far. We have thereby carried out the study for the [110]-oriented single crystal. Note that for the ferroelectric *R* phase, a non-single macroscopic domain state with two *R* twin variants can be produced by the [110] electric field. To avoid this problem, the measurements have been confined within the *HK0* scattering plane so that the in-plane momentum transfer effectively results from one single domain.

## 2. Experimental Details

0.29PIN-0.45PMN-0.26PT single crystals were grown using the vertical Bridgman method, and the nominal composition was verified using energy-dispersive X-ray spectroscopy. A disc-shaped sample (thickness/radius: ~3/20 mm; mass: ~28 g) was cut from the boule with its vertical axis oriented along the [110] crystal direction. Pt electrodes were sputtered onto the two opposing surfaces for poling the crystal under an electric field of 8 kV/cm at room temperature. Dielectric properties of the crystal over a temperature range of 100–800 K were measured on a smaller sample cut nearby using an Agilent impedance analyzer under 0.1-1000 kHz ac excitation frequencies.

Neutron scattering experiments were carried out using the TAIPAN thermal triple-axis spectrometer located at the OPAL reactor, Australian Nuclear Science and Technology Organisation (ANSTO).[24] The spectrometer was operated in a fixed final energy mode, $E_f$ = 14.87 meV. A set of double-focusing monochromator and analyzer based on the (002) Bragg reflection of highly-oriented pyrolytic graphite (HOPG) crystals were used. An HOPG filter was placed before the analyzer to remove high-order contamination energies. The horizontal beam collimation conditions were set as: *open-open*-sample-*40'-open*; *i.e.*, only one collimator was used before the analyzer. The energy resolution obtained from these configurations was about 1.1 meV, as seen from the full widths at half maximum (FWHM) of the Gaussian peaks at the elastic lines. The crystal was wrapped with Al foils, mounted on an Al sample holder and loaded into a vacuum closed-cycle cryofurnace. Below the $T_C$, the crystal was measured in both poled and unpoled (depoled) states by following an appropriate heating/cooling sequence. The crystal was aligned with the [001] axis (lying in the disc plane) vertical thus giving access to the *HK0* scattering plane. At 300 K the pseudo-cubic lattice parameter of the crystal is $a$ = 4.04 Å and thus a reciprocal lattice unit (rlu) of $2\pi/a = 1.56$ Å$^{-1}$ was set. Both constant energy (constant-$E$) scans and constant momentum (constant-$Q$) scans were made to measure the elastic diffuse scattering (i.e., $E = 0$) and phonon dispersion curves. The measured constant-$Q$ phonon spectra were fitted based on the uncoupled damped harmonic oscillator (DHO) model using the PAN/DAVE software packages.[25]

## 3. Results and Discussion

### 3.1 Low frequency dielectric behavior

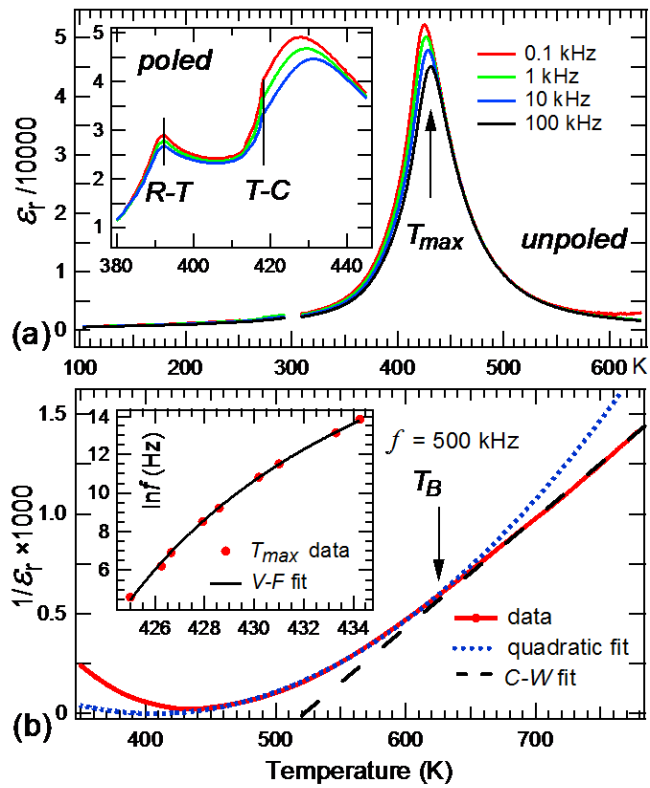

**Figure 1** (a) Temperature dependent spectra of real part dielectric permittivity ($\varepsilon'_r$) for the unpoled and (Inset) poled 0.29PIN-0.45PMN-0.26PT single crystal. For the unpoled crystal, (b) shows the inverse dielectric permittivity ($1/\varepsilon'_r$), measured at 500 kHz, as a function of temperature along with its *Curie-Weiss* and quadratic law fits; Inset shows the *Vogel-Fulcher* fit to the observed $T_{max}$ values.

Low frequency dielectric measurement is used to investigate the average structural transitions as well as the macroscopic relaxor properties of the studied crystal. Figure 1(a) shows the measured temperature dependent dielectric spectra for the unpoled crystal, and a portion of the spectra for the poled crystal (Inset) to highlight the main difference between them. Within the measured 100-800 K range, the unpoled crystal only shows a broad dielectric permittivity maximum peak at a temperature $T_{max}$ = 426.7 K (at 1 kHz). After poling, an obvious shoulder appears on the spectra around 392 K, together with a weak but clearly discernable anomaly occurring at ~418 K prior to the basically unshifted $T_{max}$. These poling induced dielectric anomalies are consistent with the results obtained from the PMN-PT system as well as [001]- and [111]-oriented PIN-PMN-PT single crystals as we recently reported.[26] On the basis of the general phase diagram of relaxor-PT systems, the anomalies at 392 K and 418 K are attributed to the $R$-$T$ and $T$-$C$ structure transitions, respectively. The monoclinic structures might be involved as intermediate phases,[27] nonetheless not considered in this scattering study because the experimental $q$ resolution is not sufficiently fine to allow identification of them.

The $T_{max}$ for both unpoled and poled crystals are dependent on the measuring frequency. Such frequency dependence is the hallmark behavior of relaxors and is usually described with respect to the *Vogel-Fulcher* formula.[28] Fig. 1(b) Inset shows the *Vogel-Fulcher* fit to the measured $T_{max}$ values for the unpoled crystal. The obtained model parameters are: the attempt frequency $f_0$ ~ 630 GHz, the activation energy $E_A$ = 304 K (~26 meV) and the freezing temperature $T_f$ = 412 K, somewhat close to the $T_C$ (= 418 K). Fig. 1(b) shows the inverse dielectric permittivity, $1/\varepsilon'_r$, measured at 500 kHz against temperature for the unpoled crystal. At this frequency, the extrinsic dielectric contribution of a finite conductivity effect at high temperatures is minimized and the intrinsic contribution from the PNRs can be reliably delineated. The $\varepsilon'_r$ follows the *Curie-Weiss* law, which determines the dielectric behavior of a displacive ferroelectric system, at a sufficiently high temperature part, as illustrated by the linear fit in Fig. 1(b). On the other hand, the high temperature side near the peak of the $1/\varepsilon'_r$ curve can be fitted to an empirical quadratic law for describing the relaxor dielectric behavior[29]:

$$\frac{1}{\varepsilon'(T)} = \frac{1}{\varepsilon_A} + \frac{(T-T_A)^2}{2\varepsilon_A \delta^2} \qquad (1)$$

where $\varepsilon_A$, $T_A$ and $\delta$ are the fitting parameters that define the peak but have no clear direct physical meaning. Both the *Curie-Weiss* and the quadratic model fits show divergence at a narrow crossover region centered at ~630 K. This transition in the temperature dependence of $\varepsilon'_r$ signifies the emergence of PNR's (*i.e.*, $T_B$ = 630 K), similar to the optical measurements.[5] The $T_B$ of the studied crystal is nearly identical to that of pure PMN obtained with the same method, and close to that (~620 K) of 0.26PIN-0.46PMN-0.28PT assessed from Brillouin spectroscopy.[30]

**3.2 Diffuse scattering in relation to the electrical poling and temperature**

Figure 2 shows the elastic diffuse scattering patterns measured near at the (100) and (110) Bragg peaks. At 300 K, the (100) and (110) patterns for the unpoled state show characteristic butterfly and ellipsoid/rod shapes, respectively, with the diffuse scattering component extending along the <110> reciprocal space directions (Figs. 2(b) and (e)). The anisotropy of the (100) pattern is somewhat weak; that is, the gaps between the <110> wings are small, compared to those observed in PMN or PZN.[7] This is due to the relatively large PT content of the studied composition and consistent with the tendency found in the PMN-*x*PT system by Matsuura *et al*.[9] After poling along the [110] direction, the (100) patterns become markedly asymmetric (Fig. 2 (a)), with one wing that is perpendicular to the poling direction greatly suppressed while the other wing parallel to the poling direction slightly enhanced. Likewise for the (110) pattern, the long axis of the ellipsoid shape appears to be truncated by the poling and a longitudinal scattering component becomes more prominent in the resultant pattern. Overall, the observed poling behavior of the diffuse scattering is similar to that of PZN-*x*PT under application of the [111] electric field as reported by Xu *et al*.[8] It may also be interpreted according to the 'pancake' model of PNR's proposed by the same authors: The local polarization inside of PNR's is along the <110> directions; the [110] field poling produces 2*R* ([111] and [11-1]) macroscopic domains in which the PNR's with local polarization directions orthogonal to the long-range ferroelectric order, *e.g.*, [1-10], are favored; on the contrary, the population of the PNR's polarized along [110] is reduced thereby leading to the suppress of the diffuse rod along the [1-10] direction in reciprocal space. Nevertheless, this interpretation should not be unique since as long as the local correlation along [110] is suppressed by the electric field the same diffuse scattering effect may occur irrespective of the specific microscopic picture of the PNR's.[31]

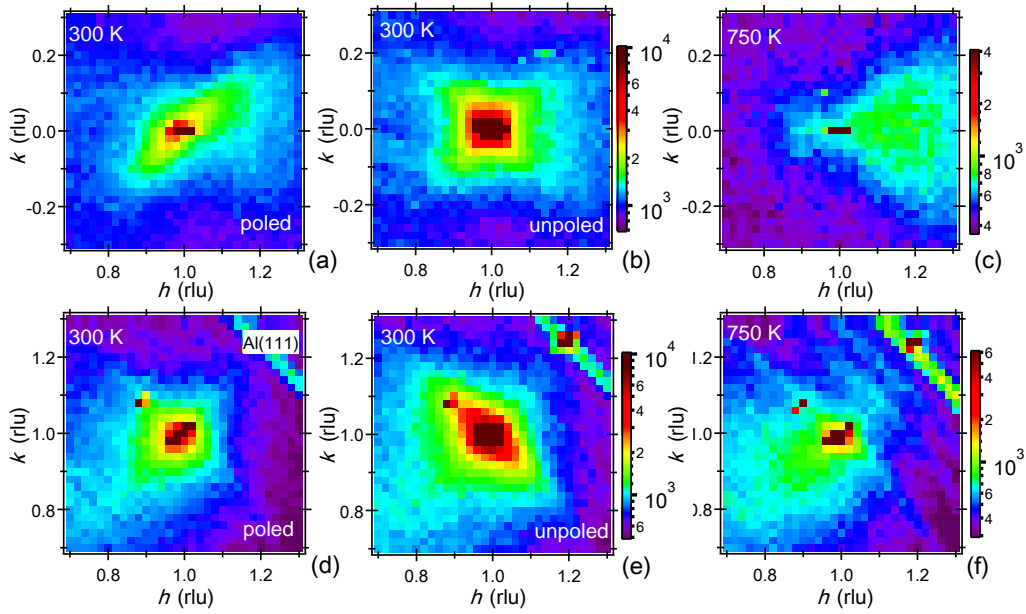

**Figure 2** Elastic diffuse scattering maps around the [(a), (b) and (c)] (100) and [(d), (e) and (f)] (110) Bragg peaks. The same color scales are used in (a)/(b) and (d)/(e). The arcs at $Q \sim 1.70$ rlu are contamination Al scattering from the sample environment.

At 750 K, the diffuse scattering intensities are much lower than at 300 K, and more significantly, the dominant diffuse component is longitudinal now thus forming distinct distribution patterns from those at 300 K (Figs. 2 (c) and (f)). As this temperature is well above the $T_B$, the PNR's population is expected to be extremely low and thus the observed diffuse scattering should point to another origin other than the PNR's. Hiraka *et al* first reported this above-$T_B$ diffuse scattering in PMN and interpreted it as an effect of the *B*-site short range chemical order.[32] Burkovsky *et al* recently attributed it to Huang scattering through quantitative theoretical modelling coupled with their measurements on PMN.[33] Huang scattering originates from elastic lattice deformations exerted by defects, *e.g.* lattice site substitutions, and appears to be a more reasonable description in our case since the *B*-site ordering in PIN-PMN-PT is presumably much weaker than PMN as a result of the mixture of four heterovalent cations. Another obvious feature on the (100) and (110) diffuse patterns at 750 K is a strong asymmetry in intensity with respect to ***Q***. For example, the high ***Q*** side of the (100) pattern has higher intensities than the low ***Q*** side. This asymmetry indicates a strong size-effect arising from the multiple *B*-site atoms which have different sizes and scattering cross sections.[34] A similar size-effect has been observed and modelled by Welberry *et al*, however, only for the low temperature PNR-related diffuse scattering in PZN.[7]

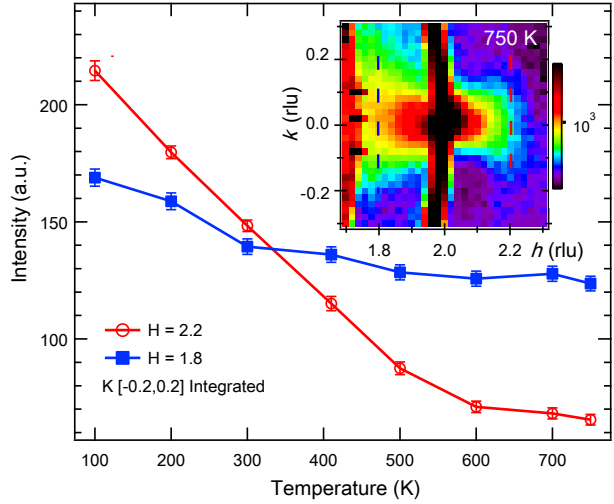

**Figure 3** Integrated diffuse scattering intensities measured from line scans (see the dash lines on Inset) near the (200) Bragg peaks as a function of temperature. Inset shows the (200) diffuse scattering map at 750 K. The two arcs on the map are contamination Al scattering, which nevertheless is well separated from the interested line scans.

Fig. 3 shows the temperature dependence of the integrated diffuse intensities of the line scans taken around the (200) Bragg peak. These two scanned lines are symmetric about the (200) peak, but as a result of the above-mentioned size effect (see also the (200) diffuse scattering profile at 750 K in Fig. 3 Inset), their intensities show large difference. At the high-$Q$ side, the intensity of the $H = 2.2$ lines starts to increase below 600 K reflecting the growth in population of the PNR's, in good agreement with the $T_B \sim 630$ K established from the dielectric data. Such a tendency of intensity is also true for the $H = 1.8$ lines but their growth rate is obviously much lower. At low temperatures *e.g.* 200 K, the intensities of the $H = 2.2$ lines surpass those of the $H = 1.8$ lines. These results agree with the previous reports that the PNR-related diffuse scattering cross section in the (200) zone is non-zero but rather weak,[10] being comparable to that of the above-$T_B$ Huang scattering. On the other hand, the relative intensity changes shown here may also suggest that the size-effect in the present system modulates the cross sections of these two coexisting types of diffuse scattering in opposite ways.

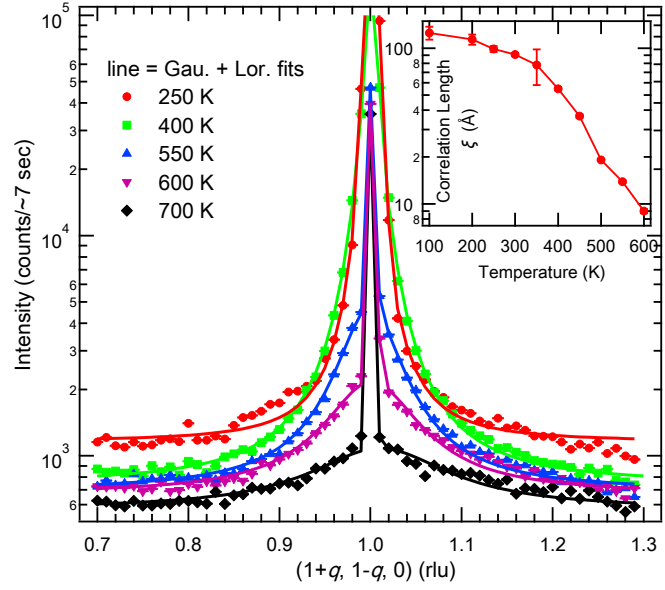

**Figure 4** Elastic diffuse scattering profiles scanned along the (1+$q$,1-$q$,0) direction at selected temperatures. There profiles are fitted with a Gaussian peak plus a Lorentzian peak. Inset shows the calculated correlation length, $\xi$, as a function of temperature.

Further line scans along the ridge of the (110) diffuse rod were made to estimate the dipole correlation length $\zeta$ (which effectively represents the PNR size) of the system. Note that the data below 500 K was collected on the unpoled state. Fig. 4 shows the scanned (1+$q$,1-$q$,0) profiles at selected temperatures along with their fits. The fitting model includes a Gaussian peak associated with the (110) Bragg peak and a Lorentzian function accounting for the PNR-related diffuse scattering. The $q$ resolution function of the spectrometer is not considered in this model. At 700 K the Lorentzian component is rather broad like a background, and it becomes narrower and more discernible as the temperature decreases below the $T_B$, followed by gradually merging with the Gaussian component. The correlation length is calculated from the inverse of the half width at half maximum of the fitted Lorentzian peaks. As shown in Fig. 4 Inset, the PNR's start to continuously grow in size at 600 K, apparently with no anomalies occurring at the intermediate temperature scales, *e.g.*, $T_{max}$ or $T_{R-C}$, and the PNR size reaches ~90 Å at 300 K. By comparison, the PNR size of the PMN-$x$PT system at 300 K was reported in Ref. [9]: $\zeta$ = 12.6 Å, 33.7 Å and 350 Å, respectively, when $x$ = 0, 0.1 and 0.2. Therefore, for the same PT content, the PNR's in PIN-PMN-PT appears to be much finer than in PMN-$x$PT.

## 3.3 Lattice dynamics measurements

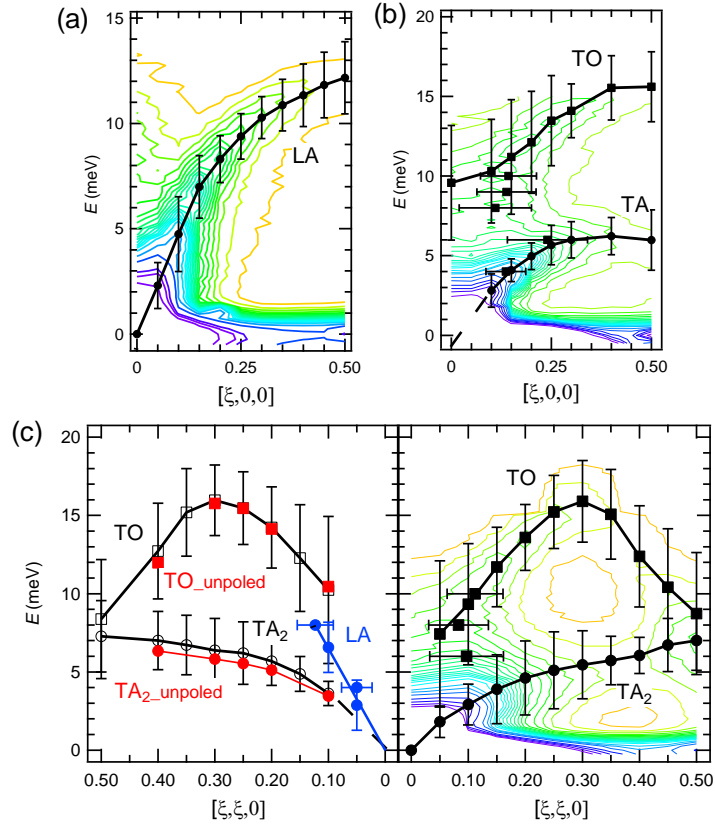

**Figure 5** Low-energy phonon dispersion curves along the [001] and [110] directions of 0.29PIN-0.45PMN-0.26PT at 300 K, all measured in the poled state except otherwise marked. The bars represent the FWHM of the fitted phonon peaks. Constant-$Q$ scan intensity contours are overlaid on some of the panels. Lines are a guide to the eyes.

Figure 5 shows the main several branches of low-energy phonon dispersion curves at 300 K studied in this work. Note that there are two non-degenerate branches of transverse acoustic phonons propagating along the [110] direction, denoted as $TA_1$[110] and $TA_2$[110] which are polarized along [001] and [1-10], respectively; the $TA_1$[110] branch is not accessible in the *HK0* scattering plane. The longitudinal acoustic (LA) branches were measured across the entire Brillouin zone along [100] while only near the zone center ($\Gamma$-point) along [110]. Overall, the LA phonons show well-defined, underdamped phonon peaks in the measured constant-$Q$ spectra. Their phonon energies have little temperature dependence and no clear anomalies related to the characteristic temperature scales are found on them. In the following text, we focus our discussion on the transverse acoustic and optical phonon behavior of the system.

Below the $T_C$ = 418 K, the TA[100] and TA$_2$[110] phonons were measured in both poled and unpoled states along the directions either perpendicular or parallel to the poling direction ***P*** with an aim to examine the electric field-induced lattice dynamical anisotropy. Due to the consideration of focusing conditions, these measurements were made in different symmetric Brillouin zones, *e.g.*, (220) and (2-20) zones. As shown in Fig. 5, the TA$_2$[110] phonon branch propagating perpendicular to the ***P*** (*i.*e., ***q*** // [1-10] and polarization vector ***ε*** // [110]) have larger energies than the same but parallel-to-the-***P*** branch (***q*** // [110] and ***ε*** // [1-10]) at 300 K. This energy difference extends from near the $\Gamma$-point to the zone boundary (0.5, 0.5, 0), as is apparent from the measured discrete phonon wave vectors $\xi$. By contrast, the TA$_2$[110] phonons measured on the unpoled state basically have the same energies along the two directions and both are located in between the two poled branches. Apart from these changes in phonon energy, the electric poling also modifies the linewidths of the TA$_2$[110] phonons and a general trend can be found that the phonons with ***q*** // [110] have broader linewidths, namely, more strongly damped. Obviously, such an acoustic phonon anisotropy cannot be due to the the macroscopic polarization effect because the TO phonons measured at the corresponding wave vectors are almost the same in both energy and linewidths (see Fig. 5(c)), though there are indeed small but discernable energy differences at the near-$\Gamma$-point wave vectors (*e.g.*, $\xi$ = 0.1), which may point to a finite rhombohedral distortion of the crystal. Similar phenomena were reported by Xu *et al* on PZN-0.045PT single crystals under the [111] electric field cooling conditions, and a strong correlation between the phonon damping behavior and the diffuse scattering cross sections leaded them to introduce the PNR-TA coupling mechanism.[20] Such a coupling mechanism is well justified in our case; the TA$_2$[110] phonons with ***q*** // [1-10] show higher energies while along that direction the diffuse intensity is greatly reduced by the poling.

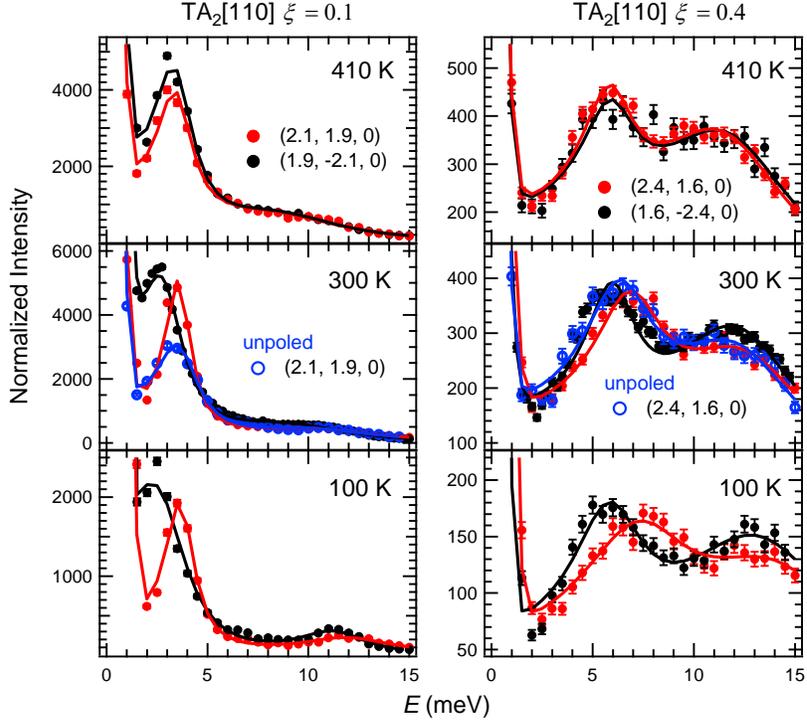

**Figure 6** Constant-$Q$ spectra of the TA$_2$[110] phonons (with reduced phonon wave vector $\xi$ = 0.1 and 0.4) measured at the (220) and (2-20) Brillouin zones, along the $q$ directions perpendicular and parallel to the poling direction, respectively. Lines are the DHO model fits to the measured spectra.

    Fig. 6 shows the constant-$Q$ scan spectra obtained at $Q$ = (2.1, 1.9, 0)/(1,9, -2.1, 0) ($\xi$ = 0.1) and $Q$ = (2.4, 1.6, 0)/(1,6, -2.4, 0) ($\xi$ = 0.4) along with their DHO model fits at selected temperatures. These spectra intuitively show the difference between the two TA$_2$[110] phonon branches. All these TA phonons do not change noticeably when the poled crystal is cooled from 300 K to 100 K, This suggests that the degree of the PNR-TA coupling degree is weakly affected by the cooling below 300 K, despite the fact that the PNR's have been found to continuously increase in size and population during the course. It has been argued by Stock *et al* that the TA-PNR coupling is dependent on their relative length scales and becomes weakened in the long-wavelength limit.[21] In the studied PIN-PMN-PT system, the PNR size at 300 K is ~90 Å, already exceeding that of $\xi$ = 0.1 ($q$ = 0.14 rlu; real space wavelength ~30 Å). Therefore a further increase of the PNR size might only affect those near $\Gamma$-point TA phonons, which nevertheless are not measurable in the

present experiment. At 410 K, a temperature intermediate between $T_{R-T}$ and $T_C$, the difference between the two TA$_2$[110] phonon branches vanishes (see Fig. 6) and so does the asymmetry of the diffuse scattering pattern (data not shown). It seems that the PNR distribution has changed drastically across this *R-T* average structure transition though the poled crystal is expected to remain a partially polarized state for the *T* phase.

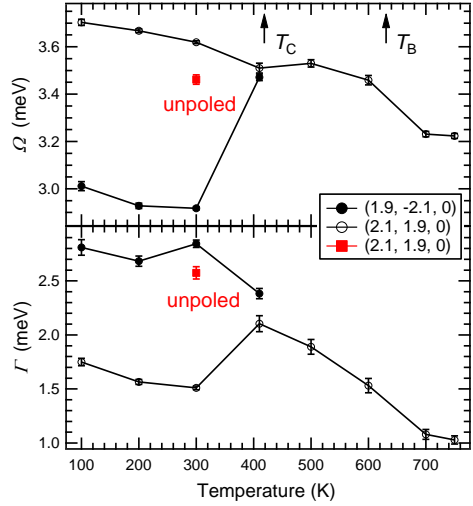

**Figure 7** Temperature dependence of the phonon energy $\Omega$ and width $\Gamma$ of the TA$_2$[110] mode at $\xi = 0.1$, measured at the (220) and (2-20) Brillouin zones along the *q* directions perpendicular and parallel to the poling direction, respectively. Error bars represent the fitting uncertainties.

In Fig. 7 the energy ($\Omega$) and linewidth ($\Gamma$) of the TA$_2$[110] phonon at $\xi = 0.1$ is plotted against the measuring temperature. Here the contrast between the poled and unpoled crystals has been shown in the previous data and therein discussed. Over the entire temperature range, the energy of this phonon decreases with increasing temperature, which may be largely attributed to the lattice anharmonicity. Note, however, that there is a noticeable anomaly for this TA$_2$[110] phonon around the $T_B$ where its energy jumps by ~7% and below the $T_B$ its linewidth varies more rapidly than above the $T_B$. In contrast to the slow varying tendency due to the anharmonic effect, this anomaly is very likely to be associated with the appearance of the PNR's. As the linear slopes of the near-$\Gamma$ portions of acoustic dispersion curves are scaled to the macroscopic elastic constants, the energy positions of the TA$_2$[110] phonon at $\xi = 0.1$ reflect the evolution of the difference of ($C_{11}$- $C_{12}$) and the jump at the $T_B$ corresponds to a 15% change of the elastic constant. Note that this observed stiffening effect (with the temperature falls through the $T_B$) is quite opposite to the trends seen in both the Brillonin scattering (probing in the GHz range) and ultrasonic measurements (in the sub-MHz) of PIN-PMN-PT and other relaxors.[30,35] We believe that this discrepancy lies in the difference in the probing frequencies and length scales of these techniques and further suggests the complexity of the PNR-acoustic coupling effect as a result of the temperature-dependent relative dynamic ranges and length scales between the PNR's and the average crystal lattice.

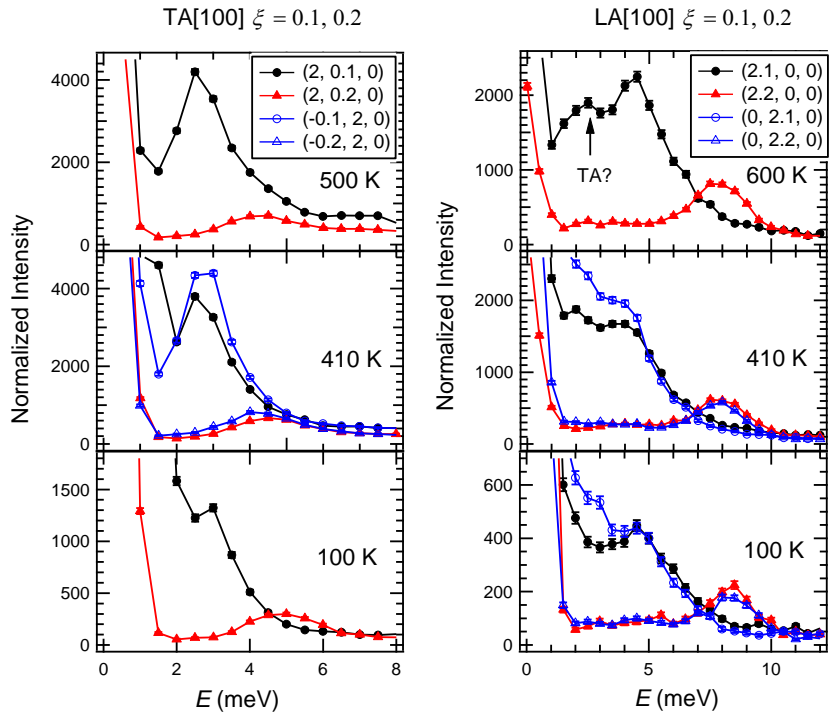

**Figure 8** Constant-$Q$ spectra of the TA[100] and LA[100] phonons ($\xi$ = 0.1 and 0.2) measured at the (200) and (020) Brillouin zones.

The TA/LA[100] acoustic branches were also examined for their potential coupling effect with the PNR's. Fig. 8 shows the constant-$Q$ spectra measured at two wave vectors, $\xi$ = 0.1 and 0.2, at selected temperatures. In the $R$ phase, the TA[100] phonons measured at the (200)/(020) zones, *i.e.*, propagating along the [010]/[100] directions, are equivalent in terms of their orientation with the poling direction. Thus no anisotropy is expected for them and similarly for the LA[100] phonons below the $T_{R-T}$, as has been confirmed by the measurements. On the other hand, as is evident from Fig. 8, the measured line shape of the TA[100] phonon at $\xi = 0.1$ are quite different at 100 K, 410 K and 500 K, whereas those phonons at $\xi = 0.2$ show little variations in line shape at the same temperatures. The TA[100] phonon at $\xi = 0.1$ is overdamped (*i.e.*, $\Gamma \geq \Omega$) at 100 K and it becomes underdamped at 500 K; at 410 K, the phonon observed at $Q = (2, 0.1, 0)$ has an intermediately damped profile which is slightly different from that at $Q = (0.1, 2, 0)$, suggesting a weak lattice anisotropy in the *T* phase. The contrast between the TA[100] phonons at two wave vectors, $\xi = 0.1$ and 0.2, agrees with a recent report by Stock *et al* that the damping of the TA[100] phonons of PMN is only observed within a limited momentum transfer range.[21] In our case, the line shape changes of the TA[100] phonon can be again explained according to the PNR size of the system: the PNR size is ~35 Å at 500 K, close to the real space wavelength of the phonon at $\xi = 0.1$, and therefore the damping of the TA[100] phonon is much weakened at this temperature. For the LA phonons, their constant-$Q$ spectra show complex line shapes at $\xi = 0.1$ but at $\xi = 0.2$ the phonon peaks are well-defined. It appears highly possible that the TA phonon scattering component leaks into the measured LA spectra near the $\Gamma$-point.[36] In this regard, the LA[100] phonon at $\xi = 0.1$ shows consistent variations in line shape with the TA phonons as a function of temperature, but it is rather difficult to ascertain a PNR-LA coupling effect based on our data.

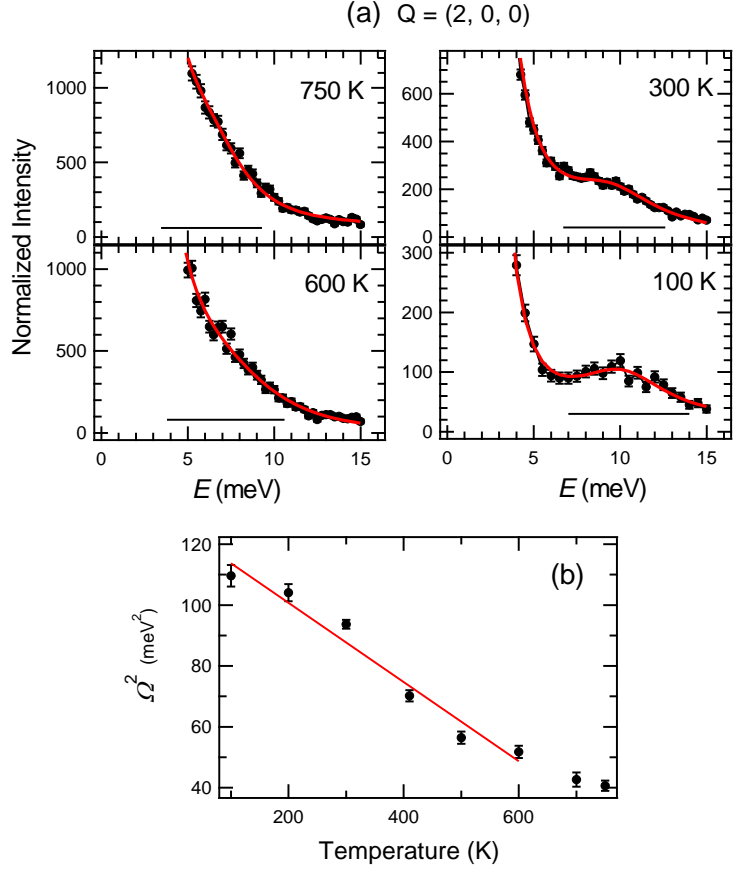

**Figure 9** (a) Constant-$Q$ spectra scanned at $Q$ = (2, 0, 0) at selected temperatures. Lines are the DHO fits to these measured spectra and horizontal bars represent the fitted phonon widths $\Gamma$. (b) The squared zone-center TO phonon energy as a function of temperature. The error bars show the fitting uncertainties.

The TO phonons of the studied PIN-PMN-PT single crystals show the waterfall features.[14,15] As shown in Figs. 5 (b) and (c), both the TO[100] and TO[110] branches have a strongly damped dispersion near at the zone center; that is, the phonon widths fitted from the constant-$Q$ spectra are very large for the low-$q$ portions and by contrast the constant-$E$ scans at several energies ($E$ = 5-10 meV) are centered at a similar wave vector, $q_{wf}$ ~ 0.15 rlu, for the both branches. Note that the waterfall phenomenon becomes more manifest at high temperatures close to the $T_B$ where the near zone-center TO phonons are overdamped. Fig. 9(a) shows the line shape evolution of the lowest $\Gamma$-point TO phonon at selected temperatures together with the DHO fits to these phonons. These results clearly demonstrate that as the temperature falls below the $T_B$ the TO phonons gradually recover an underdamped line shape with the energies shifting to a higher position. This observed behavior is suggestive of soft mode characteristics of the system. In canonical displacive systems, *e.g.*, PT, the development of ferroelectric orders is linked with the softening and condensing (namely, reaching zero energy) of a TO phonon mode that designates the ferroelectric distortions, and the soft TO mode recovers after the phase transition.[16] In Fig. 9(b), the square of the $\varGamma$-point soft TO mode ($\varOmega^2$) is plotted against temperature. The $\varOmega^2$ increases linearly with decreasing temperature below 600 K, in agreement with the soft mode picture. Above the $T_B$, the energy of this soft TO mode does not change noticeably up to 750 K, and no measurements were attempted at higher temperatures to observe the overall softening trend of this TO mode. Nevertheless, the current data does show an incomplete softening of the $\varGamma$-point TO mode, similar to that observed on PMN-$x$PT ($x$ = 0, 0.32 and 0.64).[37] The formation mechanism of PNR's in relaxors remains a highly debatable issue to date, but clearly, the picture that the soft TO phonons directly condense into PNR's is not fully justified by the above observations. It appears that an order-disorder type mechanism may be incorporated into the soft mode picture in order to give a more accurate description of the PNR dynamics of relaxors.[38]

### 4. Conclusion

We have studied the local structure and lattice dynamics of [110]-oriented 0.26PIN-0.46PMN-0.28PT single crystals in relation to the poling states and variable temperatures using elastic diffuse and inelastic neutron scattering. As reflected in the diffuse scattering measurements, the PNR's in this system start to grow continuously in size and population from the $T_B$ ~630 K, in agreement with the macroscopic dielectric behavior. At 300 K, the diffuse scattering around the (100) and (110) Bragg peaks show characteristic "butterfly" and ellipsoid shaped distribution patterns, respectively, on the *HK0* reciprocal space plane. Below the $T_C$, the distribution of the PNR-related diffuse scattering can be modified by the electrical poling indicating the response of the PNR's to an external electric field. This electric field effect also manifests a lattice dynamical anisotropy, mediated by the PNR-TA coupling mechanism, in the TA$_2$[110] phonon branches which show significant difference in the phonon energy and line width along the perpendicular/parallel-to-the-field directions. The PNR-TA coupling has also been observed on another TA phonon branch, TA[100], though only at small wave vectors. Furthermore, the temperature evolution of the PNR-TA coupling has been correlated with the length scale of the underlying PNR's. The "waterfall" phenomenon is observed on the [001] and [110] TO phonon branches, starting from a wave vector of ~0.15 rlu to the zone center. We have also observed a lowest-energy, soft TO phonon mode at the zone center, which hardens as the crystal cools below the $T_B$. This observation suggests soft mode characteristics of the present relaxor system.

**Acknowledgement**

QL, YL, RLW and JW acknowledge the support of the Australian Research Council (ARC) in the form of ARC Discovery Grants. YL also acknowledges support from the ARC Future Fellowships Program.

**Reference**

1. S.-E. Park and T. R. Shrout, J. Appl. Phys. 82, 1804 (1997)
2. S. Zhang, and F. Li, J. Appl. Phys. 111, 031301 (2012)
3. R. A. Cowley, S. N. Gvasaliya, S. G. Lushnikov, B. Roessli and G. M. Rotaru, Adv. Phys. 60, 229 (2011)
4. P. M. Gehring, J. Adv. Dielectr, 2, 1241005 (2012)
5. G. Burns and F. Dacol, Solid State Commun. 48, 853 (1983).
6. G. Xu, Z. Zhong, H. Hiraka, and G. Shirane, Phys. Rev. B 70, 174109 (2004)
7. T. R. Welberry, M. J. Gutmann, H. Woo, D. J. Goossens, G. Y. Xu, C. Stock, W. Chen and Z. G. Ye, J. Appl.Cryst. 38, 639 (2005)
8. G. Xu, J. Wen, C. Stock and P. M. Gehring, Nature Mater. 7, 562 (2006)
9. M. Matsuura, K. Hirota, P. M. Gehring, Z.-G. Ye, W. Chen and G. Shirane, Phys. Rev. B 74, 144107 (2006)
10. P. M. Gehring, H. Hiraka, C. Stock, S.-H. Lee, W. Chen, Z.-G. Ye, S. B. Vakhrushev and Z. Chowdhuri, Phys. Rev. B 79, 224109 (2009)
11. C. Stock, L. V. Eijck, P. Fouquet, M. Maccarini, P. M. Gehring, G. Xu, H. Luo, X. Zhao, J.-F. Li and D. Viehland, Phys. Rev. B 81, 144127 (2010)
12. J. Hlinka, J. Adv. Dielect., 02, 1241006 (2012)
13. A. Naberezhnov, S. Vakhrushev, B. Dorner, D. Strauch and H. Moudden, Eur. Phys. J.

B 11, 13 (1999)

14. P. M. Gehring, S.-E. Park, and G. Shirane, Phys. Rev. Lett. 84, 5216 (2000)
15. P. M. Gehring, S. Wakimoto, Z.-G. Ye, and G. Shirane, Phys. Rev. Lett. 87, 277601 (2001)
16. G. Shirane, J. D. Axe, J. Harada, and J. P. Remeika, Phys. Rev. B 2, 155 (1970)
17. I. Tomeno, J. A. Fernandez-CBaca, K. J. Marty, K. Oka, and Y. Tsunoda, Phys. Rev. B 86, 134306 (2013)
18. C. Stock, D. Ellis, I. P. Swainson, G. Xu, H. Hiraka, Z. Zhong, H. Luo, X. Zhao, D. Viehland, R. J. Birgeneau, and G. Shirane, Phys. Rev. B 73, 064107 (2006)
19. C. Stock, H. Luo, D. Viehland, J. F. Li, I. P. Swainson, R. J. Birgeneau and G. Shirane, J. Phys. Soc. Jpn. 74, 3002 (2005)
20. G. Xu, Z. Zhong, Y. Bing, Z.-G. Ye and G. Shirane, Nature Mater. 5, 134 (2008)
21. C. Stock, P. M. Gehring, H. Hiraka, I. Swainson, G. Xu, Z.-G. Ye, H. Luo, J.-F. Li and D. Viehland, Phys. Rev. B 86, 104108 (2012)
22. K. Ohwada, K. Hirota, H. Terauchi, H. Ohwa and N. Yasuda, J. Phys. Soc. Jpn. 80, 024606 (2006)
23. K. Ohwada, K. Hirota, H. Terauchi, T. Fukuda, S. Tsutsui, A. Q. Baron, J. Mizuki, H. Ohwa and N. Yasuda, Phys. Rev. B 77, 094136 (2008)
24. S. A. Danilkin, G. Horton, R. Moore, G. Braoudakis and M. E. Hagen, J. Neutron Res. 15, 55 (2007)
25. R. T. Azuah, L. R. Kneller, Y. Qiu, P. L. Tregenna-Piggott, C. M. Brown, J. R. Copley and R. M. Dimeo, J. Res. Natl. Inst. Stan. Technol. 114, 341 (2009)
26. Q. Li, Y. Liu, J. Wang, A. J. Studer, R. L. Withers, Z. Li, and Z. Xu, J. Appl. Phys. 113, 154104 (2013)
27. B. Noheda, D.E. Cox, G. Shirane, S.-E. Park, L. E. Cross, and Z. Zhong, Phys. Rev. Lett. 86, 3891 (2001)
28. D. Viehland, S. J. Jang, L. E. Cross and M. Wuttig, J. Appl. Phys. 68, 2916 (1990)
29. A. A. Bokov, and Z. –G. Ye, J. Mater. Sci. 41, 31 (2006)
30. T. H. Kim, S. Kojima and J.-H. Ko, J. Appl. Phys. 111, 054103 (2012)
31. M. Paściak, M. Wołcyrz, and A. Pietraszko, Phys. Rev. B 76, 014117 (2007)
32. H. Hiraka, S.-H. Lee, P. M. Gehring, G. Xu, and G. Shirane, Phys. Rev. B 70, 184105

(2004)

33. R. G. Burkovsky, A. V. Filimonov, A. I. Rdskoy, K. Hirota, M. Matsuura and S. B. Vakhrushev, Phys. Rev. B 85, 094108 (2012)

34. *Diffuse X-Ray Scattering and Models of Disorder*, T. R. Welberry (Oxford University Press, 2004)

35. M. A. Carpenter, J. F. Bryson, G. Catalan, S. J. Zhang and N. J. Donnelly, J. Phys. Condens, Matter. 24, 045902 (2012)

36. *Neutron Scattering with a Triple-axis Spectrometer Basic Techniques*, G. Shirane, S. M. Shapiro, and J. M. Tranquada (Cambridge University Press, 2002)

37. H. Cao, C. Stock, G. Xu, P. M. Gehring, J. F. Li, and D. Viehland, Phys. Rev. B 78, 104103 (2008)

38. M. Matsuura, H. Hiraka, K. Yamada and K. Hirota, J. Phys. Soc. Jpn. 80, 104601 (2011)